%% file: main.tex
\documentclass[11pt]{article}
 
\RequirePackage[a4paper]{geometry}
\usepackage[colorlinks,
            linkcolor=blue,
            anchorcolor=blue,
            citecolor=blue]{hyperref}
            
\usepackage{caption}
\usepackage{amsmath}
\usepackage{subfigure}
\usepackage{diagbox}
\usepackage{lscape,epsfig,graphicx,multirow}
\usepackage{indentfirst}
\usepackage[section]{placeins}
\usepackage{makecell}
\usepackage{color}
\usepackage{tabularx}
\usepackage[affil-it]{authblk}
{\normalsize {\large {\Large }}}
\title{Research of radon diffusion behavior in liquid scintillator}
\input{author}
\date{}

\begin{document}
\maketitle
 
\begin{abstract}

Radon and its daughters are one of the most important background sources for low-background liquid scintillator (LS) detectors. The study of the diffusion behavior of radon in the LS contributes to the analysis of the related background caused by radon. Methodologies and devices for measuring radon's diffusion coefficient and solubility in materials are developed and described. The radon diffusion coefficient of the LS was measured for the first time and the solubility coefficient was also obtained. In addition, the radon diffusion coefficient of the polyolefine film which is consistent with data in the literature was measured to verify the reliability of the diffusion device.

\end{abstract}

\begin{flushleft}
\end{flushleft}


\section{Introduction}

In order to find rare physical events of interest, the sensitive volume of the detector is gradually increased, and the background requirements of the detector are becoming increasingly stringent in the field of particle physics experiments. Liquid scintillator (LS) is widely used as targets in large detectors with excellent performance, such as JUNO~\cite{JUNO}, Borexino~\cite{Borexino}, KamLand~\cite{KamLAND}. In order to meet the requirements of physical experiments, it is necessary to reduce the background of LS and strictly control the background source. 

Natural radioactive radon dissolved in LS and its natural radioactive daughters, especially $^{210}$Pb with a half-life of 22.3 years, will be one of the most important background sources which decide whether the experiment is successful or not. There are various sources of radon, such as air leakage, cover gas, and emanating from detector materials, which can diffuse into LS. The distribution of radon caused by the radon migration properties including the diffusion and solubility in LS needs to be studied to provide a reliable basis for future experimental analysis of the background caused by radon and lead. The radon isotopes that have an impact on the background are mainly $^{222}$Rn and $^{220}$Rn. Although they have different decay lifetimes and daughters, their diffusion properties should be the same. There is no measurement result of the radon diffusion coefficient of LS in the existing literature~\cite{MUJAHID2005106, WOJCIK19918, ARAFA2002207}. 

There are generally steady-state methods and unsteady-state methods for measuring the radon diffusion coefficient of other materials~\cite{twomethod, steady-state}. In this paper, the unsteady-state method will be used to measure the radon diffusion coefficient of LS. Compared with the steady-state method, the former has a shorter measurement time (less than one day)~\cite{fast,Wojcik:2000vd}, while the latter usually has a measurement period of 2 to 4 weeks~\cite{steady}. A shorter measurement duration can not only improve the measurement efficiency but also improve the measurement stability and keep the initial state of the measured sample (such as sample volatilization), thus improving the measurement accuracy. However, this requires a highly sensitive radon detector, because it is necessary to monitor the continuously changing radon concentration. In the existing literature, the scintillation methods such as ZnS(Ag) are used to measure radon concentration~\cite{WOJCIK19918}, which requires a period of time (about 3 hours) for $^{222}$Rn and its progenies ($^{218}$Po, $^{214}$Po) to reach equilibrium. In this paper, the SIN-Pin detector is used for the first time to measure the concentration of radon diffusion, which can directly measure the first generation decay daughter $^{218}$Po of $^{222}$Rn, thus improving the sensitivity to radon concentration change.

\section{Measurement principle}

\begin{figure}[htbp]
\centering
\includegraphics[width=0.5\textwidth]{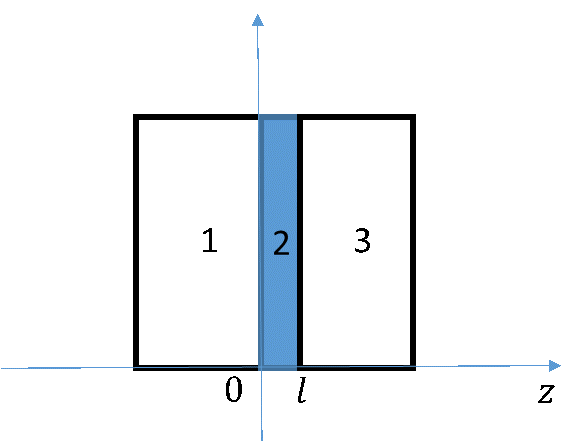}
\caption{\label{fig:principle} Schematic diagram of diffusion coefficient measurement principle.}
\end{figure}

Generally, the device for measuring the diffusion coefficient of $^{222}$Rn in a certain material can be simplified as shown in Fig.~\ref{fig:principle}, where 2 indicates the measured material, 1 indicates the $^{222}$Rn source chamber containing a constant $^{222}$Rn concentration, and 3 indicates the collection chamber. When the measured material with certain thickness contact with the gas with constant $^{222}$Rn concentration, $^{222}$Rn will first dissolve into the material surface. Because there is a $^{222}$Rn concentration gradient in the material, $^{222}$Rn will diffuse and decay in the material, and the $^{222}$Rn concentration at a certain point in the material will change with position and time. This process is usually called unstable diffusion. After a long time of diffusion and decay equilibrium, the concentration at each point does not change, and the whole system is in a stable diffusion process. The concentration in chamber 1 is C$_{0}$, the concentration in chamber 3 is C$_{1}$. The unstable diffusion process of $^{222}$Rn in the material can be described by Fick's second law:
\begin{equation}
    \frac{\partial C(z,t)}{\partial t} = D \frac{\partial^{2} C(z,t)}{\partial^{2} z} - \lambda C(z,t)
\end{equation}
where C(z,t) is the $^{222}$Rn concentration at a certain point in the material, D is the diffusion coefficient of $^{222}$Rn in the measured material, $\lambda$ is the decay constant of $^{222}$Rn. The number of $^{222}$Rn atoms in the chamber 3, y(t), can be described by Eq.(1)~\cite{Wojcik:2000vd}:
\begin{equation}
    y(t) = \frac{A D S C_0}{H \lambda} [1 - e^{-\lambda t} + 2  \sum_{n=1}^\infty (-1)^n  \times \frac{\lambda}{\beta} (1 - e^{- \beta t})]
\end{equation}
where $\beta = (D n^2 \pi^2 / H^2) + \lambda$. S is the $^{222}$Rn solubility coefficient for the measured material, A is the measured material surface area (through which Rn diffuses), and H is its thickness.


After $^{222}$Rn and its decay daughters have reached equilibrium, their radioactivity are almost equal, among which $^{218}$Po is the first generation decay daughter with a half-life of 3.1~minutes. Using the $^{218}$Po events with a short equilibrium time allows a timely response to changes caused by the diffusion behaviour of radon. Therefore, the counts of $^{218}$Po are usually used to characterise changes in the concentration of $^{222}$Rn in experiments with continuous online monitoring.

Under the condition that the radon detector keeps the same counting efficiency and detection volume, the counts from the $^{222}$Rn source and the $^{222}$Rn diffused from LS are measured respectively, and the ratio of the counting rates obtained is:
\begin{equation}
    \frac{n_2}{n_1} (t) = \frac{\eta \lambda^{'} y_2 (t)}{\eta \lambda^{'} y_1 (t)} = \frac{\eta \lambda y (t)}{\eta \lambda C_0 V_d} = \frac{y (t)}{C_0 V_d}
\end{equation}
where n$_1$ is the count rate of the detector when the $^{222}$Rn concentration inside is C$_0$; n$_2$ is the count rate corresponding to the amount of Rn diffused through the measured material; $\eta$ is the counting efficiency of the detector; $\lambda^{'}$ is the decay constant of $^{218}$Po; y$_1$(t) and y$_2$(t) are the numbers of $^{218}$Po atoms in Rn detector when measuring n$_1$ and n$_2$, respectively; V$_d$ is the volume of Rn detector.

It can be obtained from Eq.(1) and Eq.(2):
\begin{equation}
    \frac{n_2}{n_1} (t) =  \frac{A D S}{H \lambda V_d} [1 - e^{-\lambda t} + 2  \sum_{n=1}^\infty (-1)^n  \times \frac{\lambda}{\beta} (1 - e^{- \beta t})]
\label{eq:fitfunc}
\end{equation}
According to Eq.~\ref{eq:fitfunc}, the least-square method is used to fit the ratio $n_2/n_1$ curve of time function, and the expected values of these two parameters are obtained.

\section{Experimental apparatus}

\subsection{Radon diffusion apparatus}

\begin{figure}[htbp]
\centering
\includegraphics[width=0.8\textwidth]{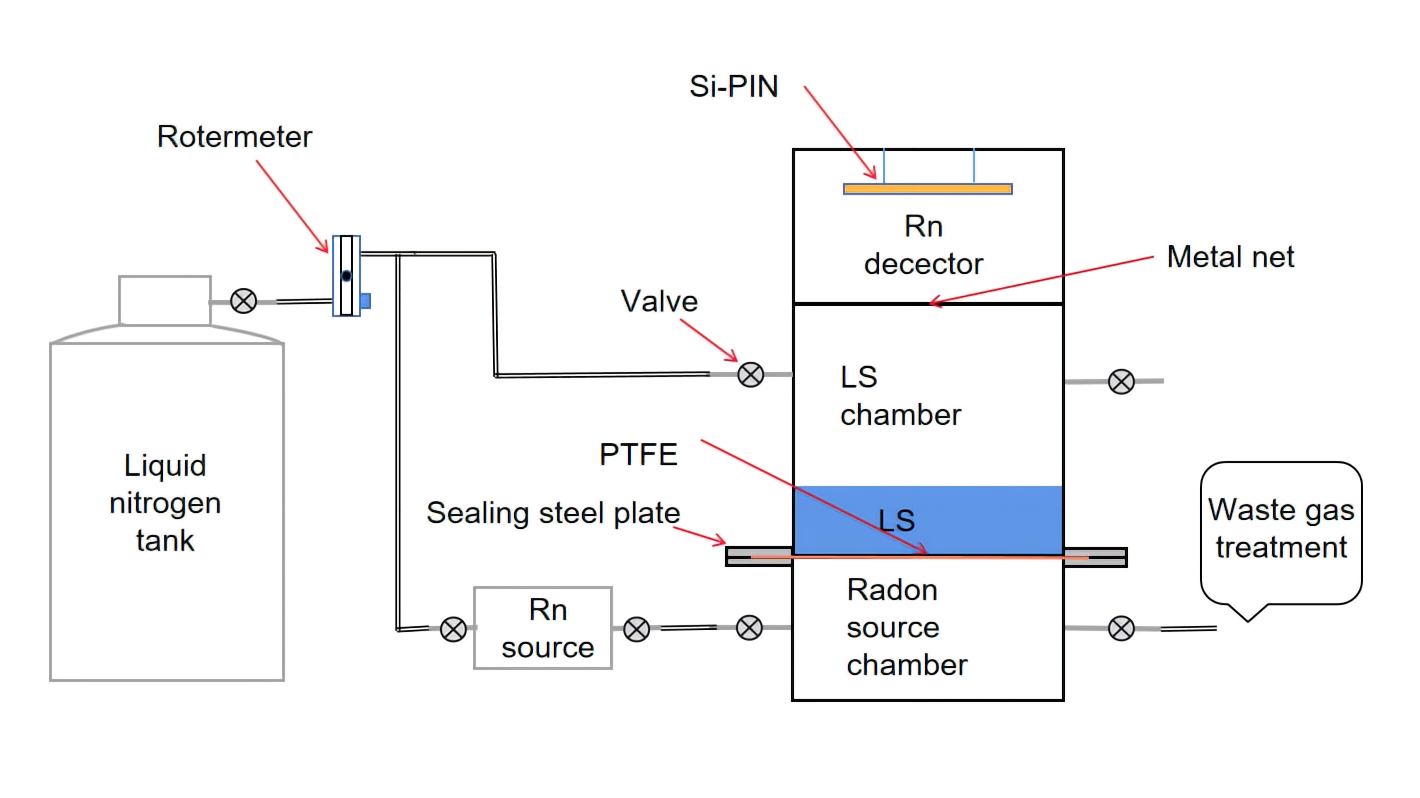}
\caption{\label{fig:Diffusion detector} The schematic diagram of the apparatus for the measurement of diffusion and solubility coefficient in LS.}
\end{figure}

The schematic diagram of the radon diffusion apparatus is shown in Fig.~\ref{fig:Diffusion detector}. The apparatus mainly consists of radon source chamber, LS chamber, radon detector and some auxiliary devices such as liquid nitrogen tank, radon source, flow meter and so on. The radon source chamber, LS chamber and radon detector are all made of stainless steel with an inner diameter of 5~cm, where the radon source chamber, LS chamber and radon detection chamber are 6.2~cm, 9~cm and 4.4~cm high respectively. The flange contact surfaces of the radon source chamber and the LS chamber are sealed by three concentric nylon rubber rings to prevent leakage. The oleophobic PTFE membrane used to carry the LS is placed between rubber rings. In order to reduce the influence of the membrane on the radon diffusion process, i.e. to ensure that the airflow can pass unhindered without the LS penetrating. After several material selections and experiments, the membrane was finally characterised as a permeable film with 5~$\mu$m pore size and a thickness of about 1~mm. The LS is placed on the upper surface of the membrane by means of a pipette gun with an accuracy of 1~ml and the thickness of the LS is controlled according to the volume of LS injected and the diameter of the chamber. 

The radon after diffusion through the LS is counted by a radon detector. The principle of the radon detector is to measure the alpha decay of electrostatically collected radon daughters to measure radon counts by creating an electrostatic field in the chamber based on the fact that the vast majority of radon daughters are positively charged~\cite{Chen2022, Radon-daughters}. To ensure that the electrostatic field does not have an influence on the diffusion behaviour of radon in the LS, a metal mesh is placed between the radon detector and the LS chamber to act as an electrostatic shield. 

The radon ($^{222}$Rn) source is a flow-through radon source made from BaRa(CO$_2$)$_3$ powder from the Radon Laboratory of the University of South China, which produces radon concentration that is dependent on the gas flow rate. During the experiment the flow rate is set at 0.5~L/min and the radon concentration is about hundred thousand~Bq/m$^{3}$. At the same time, to ensure that the pressure of evaporated nitrogen coming out of the liquid nitrogen tank is stable, a pressure stabilisation valve and a flow stabilisation valve are installed at the outlet of the tank. Before each experimental measurement, the gas in the chamber needs to be replaced with evaporated nitrogen to exclude the effects of radon and humidity in the air. The whole apparatus is placed in a room with a temperature of approximately (26$\pm$1) $^{\circ}$C.

\subsection{Detector performance}

\begin{figure}[htbp]
\centering
\subfigure[The waveforms of $^{218}$Po and $^{214}$Po signal.]{\includegraphics[width=0.45\textwidth]{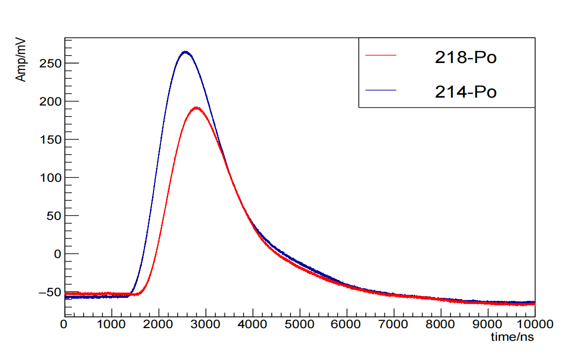}}
\subfigure[The amplitude spectrum of the waveform of $^{218}$Po and $^{214}$Po signal.]{\includegraphics[width=0.45\textwidth]{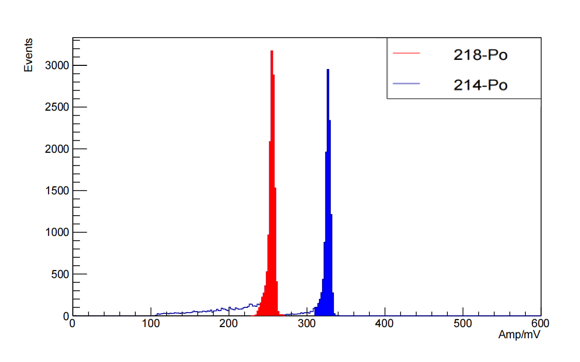}}
\caption{An example waveform of $^{218}$Po and $^{214}$Po is shown in (a) and the maximum amplitude distributions of $^{218}$Po and $^{214}$Po is shown in (b).}
\label{fig: Po}
\end{figure}

 The signal from the radon detector can be output using preamplifiers, mains amplifiers and oscilloscopes and other electronics~\cite{Sin-PIN}. As shown in Fig.~\ref{fig: Po} (a), An example waveform for $^{218}$Po and $^{214}$Po signal is shown in Fig.~\ref{fig: Po}(a). From the maximum amplitude of each signal waveform, the amplitude spectrum can be obtained as shown in Fig.~\ref{fig: Po}(b). According the the amplitude spectrum, the total counts of $^{218}$Po or $^{214}$Po can be obtained. In the following, where not specifically noted, count rates refer to counts in $^{218}$Po due to the shorter life time.

\begin{figure}[htbp]
\centering
\includegraphics[width=0.6\textwidth]{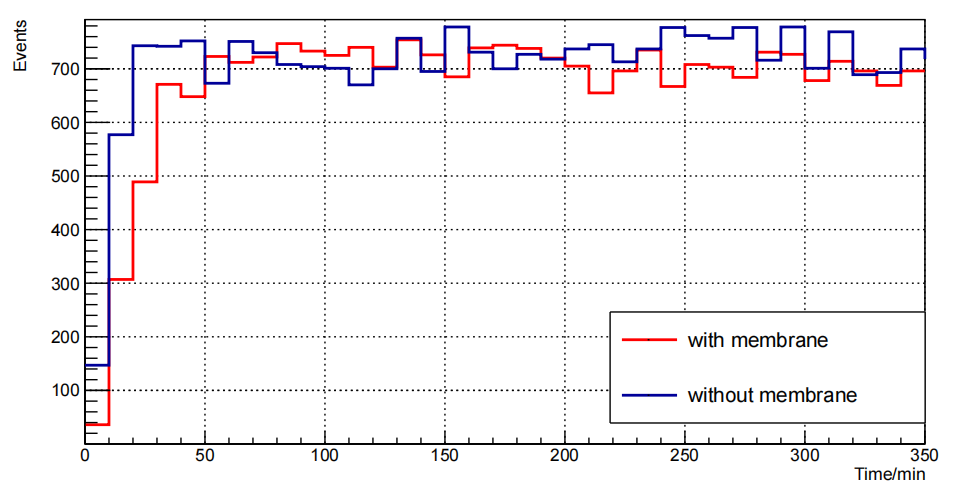}
\caption{\label{fig:membrane} The comparison of count rates with and without membrane.}
\end{figure}

Considering that radon also has a diffusion behaviour in the membrane carrying the LS, the selection of the membrane for the radon diffusion apparatus requires that the effect of radon diffusion in the membrane is minimised, i.e. similar to using the membrane with the maximum diffusion coefficient. To achieve this, a membrane with a microporous structure that is gas permeable but oleophobic was chosen. The comparison of count rates with and without membrane in Fig.~\ref{fig:membrane} illustrates that diffusion equilibrium was reached after 50 minutes. In order to reduce the influence of the membrane in the measurement of radon diffusion coefficients in the LS, the data obtained for measuring the diffusion coefficient of radon in LS after 50 minutes was used to be analyzed.

\section{Results and discussions}

\subsection{Results analysis}

\begin{figure}[htbp]
\centering
\includegraphics[width=0.6\textwidth]{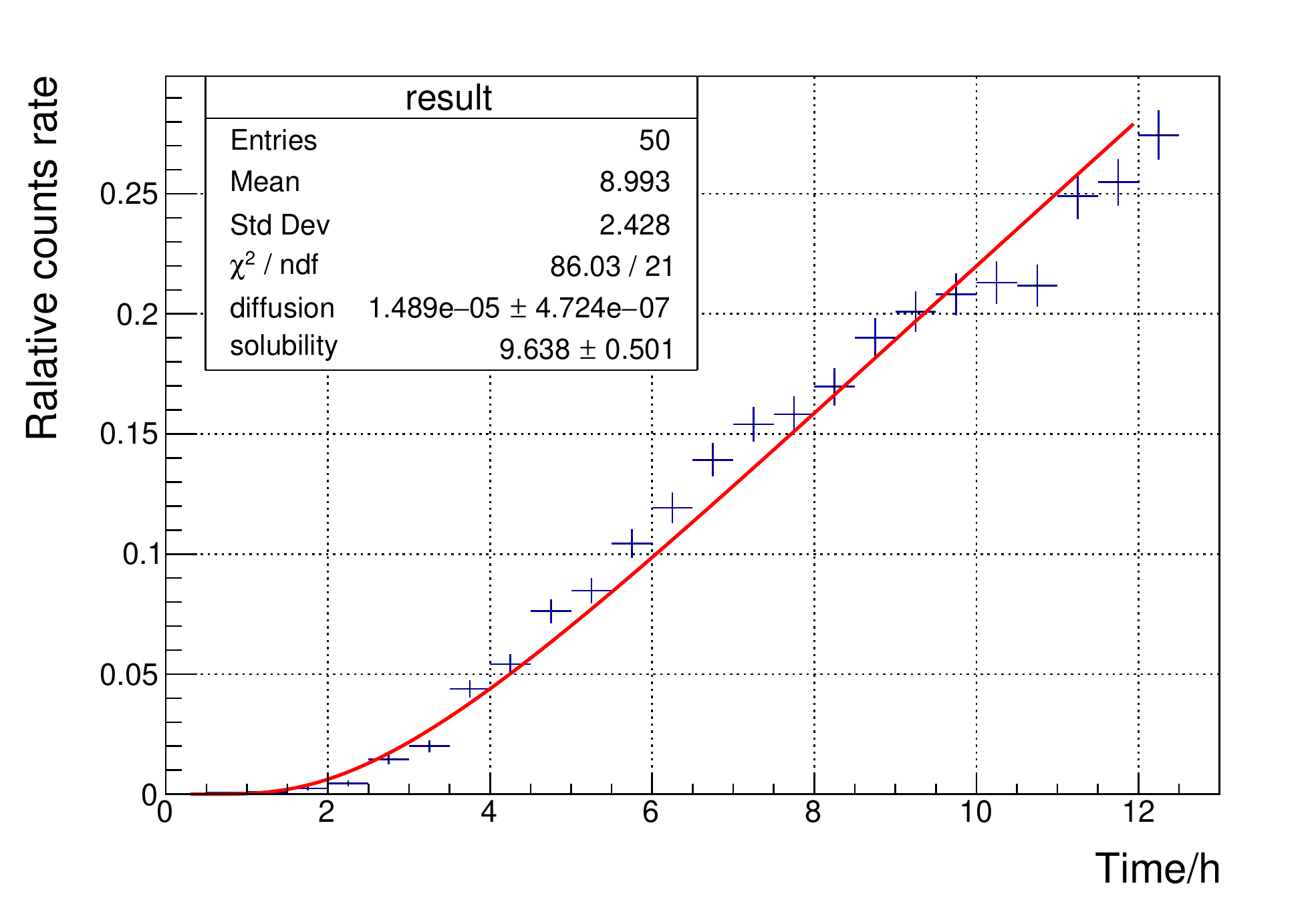}
\caption{\label{fig:rate} The dependence of the relative count rate on time for the 1~cm LS. The blue dots with error bars are experimental data and the red curve is the fitting results with the Eq.~\ref{eq:fitfunc}. The diffusion and solubility coefficients can be obtained based on the fitting results.}
\end{figure}

Based on Eq.~\ref{eq:fitfunc}, the variation in n$_{1}$ and n$_{2}$ over time needs to be measured. The n$_{1}$ can be measured by installing the membrane but not injecting LS, the results of which are shown in Fig.~\ref{fig:membrane}. The n$_{2}$ indicates the count rate measured after the injection of LS and radon diffusion within it under the same experimental conditions. The variation of n$_2$/n$_1$ with time was measured after injection of LS with a thickness of 1~cm, as shown in Fig.~\ref{fig:rate}. Using Eq.~\ref{eq:fitfunc} to fit the experimental data as shown in the red curve in Fig.\ref{fig:rate}, the diffusion coefficient (D) and solubility (S) of radon in LS can be obtained simultaneously shown in table~\ref{tab:results}. 

To ensure the reliability and stability of the apparatus as well as the measurement results, the diffusion coefficients and solubility at different thicknesses such as 0.5~cm thickness of LS are also measured, as shown in table~\ref{tab:results}. The results for the diffusion coefficients and solubility at different thicknesses of LS are consistent.

\begin{table}
\centering
\setlength{\tabcolsep}{4.5mm}{
\footnotesize
\begin{tabular}{|c|c|c|c|c|c|}
\hline
Material & Thickness & \multicolumn{2}{c|}{Measured data} & \multicolumn{2}{c|}{Literature data} \\\hline
 &  & Diffusion[$cm^2/s$] & Solubility  & Diffusion[$cm^2/s$] & Solubility\\\hline
 LS & 0.5 cm & $(1.33\pm0.10) \times 10^{-5}$ & $9.34\pm0.95$ & - & -\\\hline
 LS & 1 cm & $(1.49\pm0.05) \times 10^{-5}$ & $9.64\pm0.50$ & - & -\\\hline
 PE & 0.01 cm & $(2.00\pm1.04) \times 10^{-8}$ & $7.52\pm3.97$ & $3.7^{+2.0}_{-1.2}\times 10^{-8}$~\cite{Wolfgang} & -\\\hline
\end{tabular}}
\caption{\label{tab:results} Measurement results}
\end{table}

The radon diffusion coefficient and solubility of other membranes also can be measured, just replace the oleophobic PTFE membrane with a test membrane. As the diffusion coefficient of radon in LS is being measured for the first time, there is no literature to refer to for comparison. For this consideration, the diffusion coefficient of radon in polyethylene film (PE) used in the literature~\cite{Wolfgang} has been measured using the same apparatus and method described above. Both the experimental measurements and the results in the literature are shown in table~\ref{tab:results} and it can be seen that the results are consistent, which shows that measured diffusion coefficient the measurements of radon in LS are correct.

\subsection{Solubility measurement}

In order to verify the measured results, another novelty method is used to measure the solubility coefficient of radon in LS, and a solubility device is designed for this purpose. 

\begin{figure}[htbp]
\centering
\includegraphics[width=0.7\textwidth]{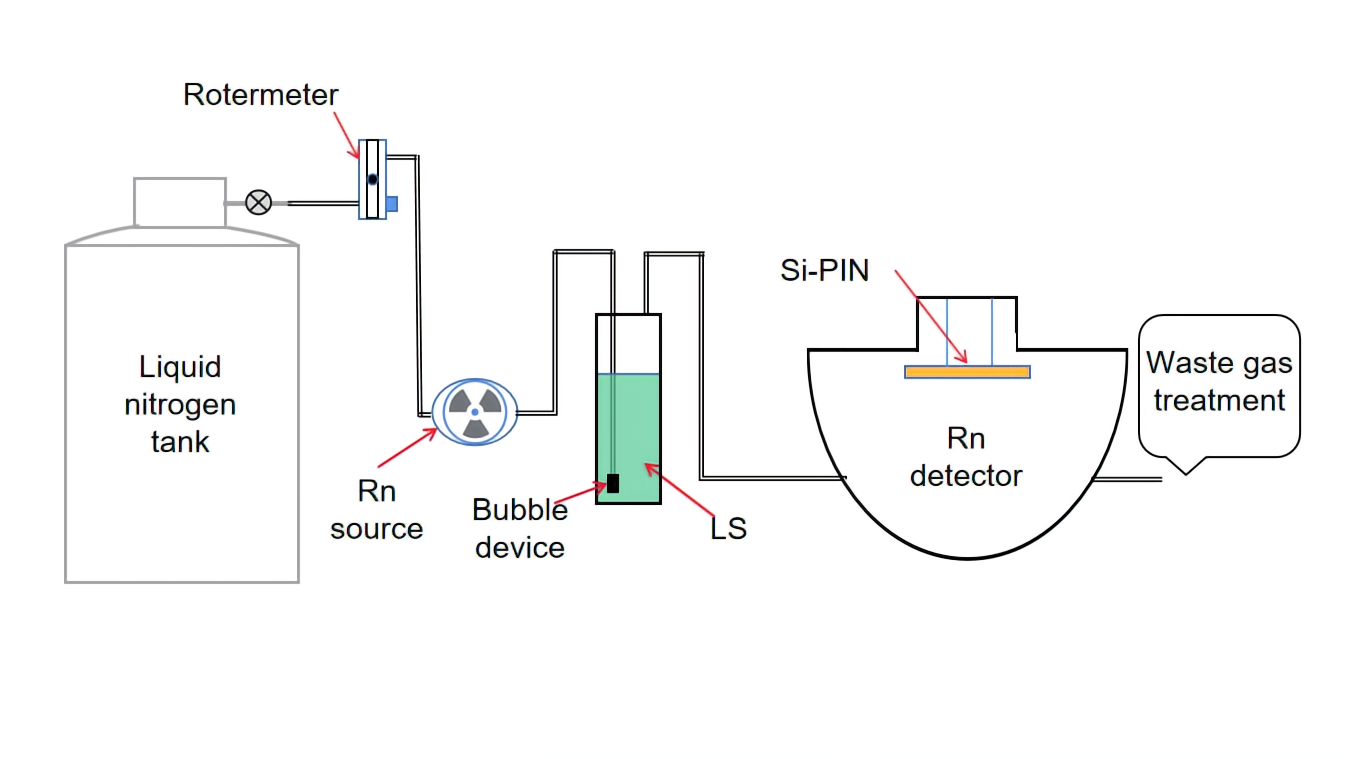}
\caption{\label{fig:Solubility detector} Solubility device.}
\end{figure}

At a certain temperature and atmospheric pressure, a gas with a stable concentration of radon is blown into the LS at a certain flow rate. When the radon in the carrier gas reaches the dissolution equilibrium with the LS, the solubility is:
\begin{equation}
    S = \frac{C_l}{C_g}
    \label{eq:solubility1}
\end{equation}
Where S is the solubility of radon in LS; $C_l$ is the radon concentration in LS and $C_g$ is the radon concentration in gas. In contrast to the methods of directly measuring radon concentrations in LS, the proposed novel method is to use a radon detector online to indirectly measure radon concentrations at the gas outlet of LS chamber.

As shown in Fig.~\ref{fig:Solubility detector}, the solubility device consists of a radon detector and a LS chamber and some auxiliary devices such as liquid nitrogen tank, radon source, flow meter and so on. The radon detector is a hemispherical chamber made of polished stainless steel for improved collection efficiency, with a volume of about 1~L. The principle of the radon detector is the same as that described in the radon diffusion apparatus. A sealed stainless steel chamber is filled with 500~ml LS and a foaming device (marked the bubble device shown in  Fig.~\ref{fig:Solubility detector}) is placed inside to improve the radon adsorption efficiency of the LS by increasing the adequate mixing of the LS with the radon carrier gas. In the whole measuring device, knife edge flanges (CF) with metal gaskets and VCR pipelines with metal gaskets are used. The leakage rate of the device is better than 1×1$0^{-9}$ Pa·$m^3$/s, which is measured by a helium leak detector (ZQJ-3000, KYKY Technology Co. Ltd).

\begin{figure}[htbp]
\centering
\includegraphics[width=0.7\textwidth]{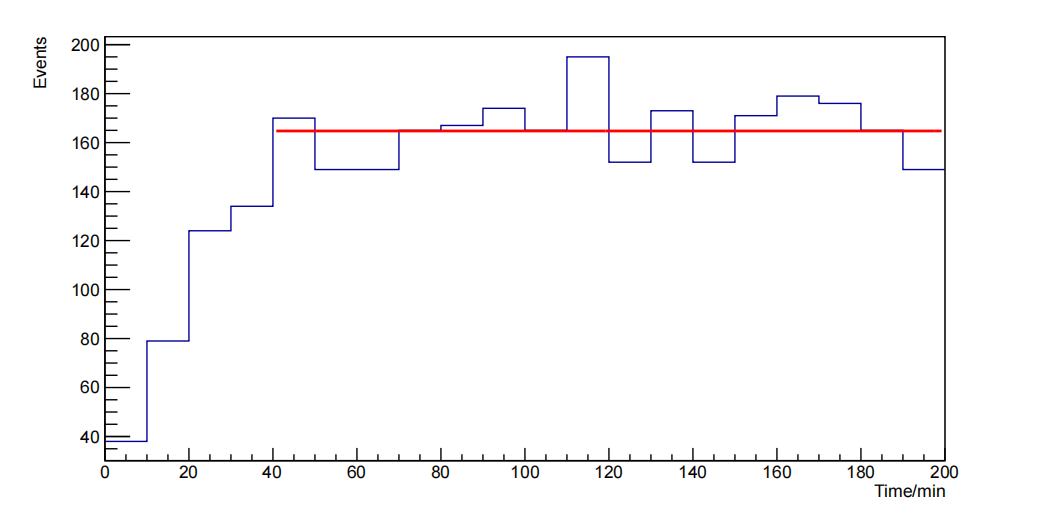}
\caption{\label{fig:S result} Variation of the count rate of the radon carrier gas after passing the LS with time, where the red line is the average of the count rate after the LS adsorption equilibrium.}
\end{figure}

The radon carrier gas with about (1600$\pm$100)~Bq/m$^3$ is introduced into the LS, and the radon detector continuously monitors the radon concentration at the gas outlet of LS chamber. The obtained count rate variation over time is shown in Fig.~\ref{fig:S result}. The counting rate continued to rise in the first 40 minutes and then stabilized in a certain range. It can be concluded that radon has reached the solubility equilibrium in LS at $\sim$40 minutes. The count rate at equilibrium also represents the count rate of the radon source itself.

According to Eq.(5), the solubility of radon in LS can be calculated according to Eq.~\ref{eq:solubility1}:
\begin{equation}
    S = \frac{\Delta n V_g}{n_1 V_l}
    \label{eq:solubility2}
\end{equation}
Where $\Delta$n is the difference of the first 40 minutes count between radon source after LS absorption and radon source; $n_1$ is the average count rate of radon source after 40 minutes; $V_g$ is the volume of radon source flowing into LS within 40 minutes, and $V_l$ is the volume of LS. The adsorption of radon in the air by the LS before it is filled into the LS chamber is not considered here. Because the concentration of radon source is much higher compared to that in the air. The temperature is stable at $(26\pm1)\,^{\circ}\mathrm{C}$ during the experiment. Based on Eq.~\ref{eq:solubility2}, the solubility can be obtained from the experimental data as $(9.6\pm2.1)$, which is consistent with the results measured by the radon diffusion apparatus in table~\ref{tab:results}.

\section{Conclusion}

By understanding the radon distribution due to diffusion in  LS, the effect of radon on the background in low-background LS detectors can be further analyzed. A dedicated device for measuring diffusion coefficients and solubility coefficients has been developed. The radon diffusion coefficient in the LS was measured for the first time and it was (1.49$\pm$0.05) $\times$ 10$^{-5}$ cm$^2$/s at about  1~cm thickness of LS. Meanwhile, the radon solubility in LS is 9.64$\pm$0.50. 

To ensure the reliability of the diffusion measurement results, two approaches were taken to verify this. One is to use this device to measure the radon diffusion coefficient of PE films and the results indicate good agreement with the literature data. The second is to adopt a new method of indirectly measuring the radon solubility by installing another device dedicated to the measurement of the radon solubility coefficient in the LS. The experimental results show that the solubility results of the two devices are self-consistent. 

The developed devices can also be used to measure radon diffusion coefficient and solubility coefficient in various materials.

\section{Acknowledgements}
This work is supported by the National Natural Science Foundation of China (Grant No. 11875280, No.11905241), and the National Natural Science Foundation of China - Yalong River Hydropower Development Co., LTD. Yalong River Joint Fund (Grant No. U1865208).

\bibliographystyle{unsrt}
\bibliography{reference}
\end{document}

%% file: author.tex
\author[a]{Zhongfang Xu}
\author[b,c,d]{Cong Guo}
\author[b,c,d]{Jincang Liu}
\author[b,c,d]{Yongpeng Zhang\thanks{ypzhang1991@ihep.ac.cn}}
\author[b,c,d]{Peng Zhang}
\author[b,c,d]{Changgen Yang}
\author[a]{Quan Tang\thanks{tangquan528@sina.com}}
\author[a]{Yu Liu}
\author[a]{Chi Li}
\author[a]{Tianyu Guan}

\affil[a]{School of Nuclear Science and Technology, University of South China, Hengyang, China}
\affil[b]{Experimental Physics Division, Institute of High Energy Physics, Chinese Academy of Sciences, Beijing, China}
\affil[c]{School of Physics, University of Chinese Academy of Sciences, Beijing, China}
\affil[d]{State Key Laboratory of Particle Detection and Electronics, Beijing, China}

%% file: main.bbl
\begin{thebibliography}{10}

\bibitem{JUNO}
Fengpeng An et~al.
\newblock {Neutrino Physics with JUNO}.
\newblock {\em J. Phys. G}, 43(3):030401, 2016.

\bibitem{Borexino}
G.~Alimonti et~al.
\newblock The borexino detector at the laboratori nazionali del gran sasso.
\newblock {\em Nucl. Instr. and Meth. A}, 600(3):568--593, 2009.

\bibitem{KamLAND}
K.~Eguchi et~al.
\newblock {First results from KamLAND: Evidence for reactor anti-neutrino
  disappearance}.
\newblock {\em Phys. Rev. Lett.}, 90:021802, 2003.

\bibitem{MUJAHID2005106}
S.A. Mujahid, S.~Hussain, A.H. Dogar, and S.~Karim.
\newblock Determination of porosity of different materials by radon diffusion.
\newblock {\em Radiation Measurements}, 40(1):106--109, 2005.

\bibitem{WOJCIK19918}
M.~Wojcik.
\newblock Measurement of radon diffusion and solubility constants in membranes.
\newblock {\em Nuclear Instruments and Methods in Physics Research Section B:
  Beam Interactions with Materials and Atoms}, 61(1):8--11, 1991.

\bibitem{ARAFA2002207}
Wafaa Arafa.
\newblock Permeability of radon-222 through some materials.
\newblock {\em Radiation Measurements}, 35(3):207--211, 2002.

\bibitem{twomethod}
Tomozo SASAKI, Yasuyoshi GUNJI, and Takeshi OKUDA.
\newblock Transient-diffusion measurement of radon in japanese soils from a
  mathematical viewpoint.
\newblock {\em Journal of Nuclear Science and Technology}, 43(7):806--810,
  2006.

\bibitem{steady-state}
Andrey Tsapalov, Loren Gulabyants, Mihail Livshits, and Konstantin Kovler.
\newblock New method and installation for rapid determination of radon
  diffusion coefficient in various materials.
\newblock {\em Journal of Environmental Radioactivity}, 130:7--14, 2014.

\bibitem{fast}
Masahiro Hosoda, Shinji Tokonami, Atsuyuki Sorimachi, Miroslaw Janik, Tetsuo
  Ishikawa, Yoshinori Yatabe, Junya Yamada, and Shigeo Uchida.
\newblock Experimental system to evaluate the effective diffusion coefficient
  of radon.
\newblock {\em Review of Scientific Instruments}, 80(1):013501, 2009.

\bibitem{Wojcik:2000vd}
M.~Wojcik, W.~Wlazlo, G.~Zuzel, and G.~Heusser.
\newblock {Radon diffusion through polymer membranes used in the solar neutrino
  experiment Borexino}.
\newblock {\em Nucl. Instrum. Meth. A}, 449:158--171, 2000.

\bibitem{steady}
Nielson~KK. Rogers~VC.
\newblock Correlations for predicting air permeabilities and $^{222}$rn
  diffusion coefficients of soils.
\newblock {\em Health Phys.}, 61(2):225--30, 1991.

\bibitem{Chen2022}
Y.Y. Chen, Y.P. Zhang, Y.~Liu, J.C. Liu, C.~Guo, P.~Zhang, S.K. Qiu~C.G. Yang,
  and Q.~Tang.
\newblock A study on the radon removal performance of low background activated
  carbon.
\newblock {\em JINST}, 17:P02003, 2022.

\bibitem{Radon-daughters}
P.~Kortrappa, S.K. Dua, P.C. Gupta, and Y.S. Mayya.
\newblock Electret - a new tool for measuring concentrations of radon and
  thoron in air.
\newblock {\em Health Phys}, 41:35--46, 1981.

\bibitem{Sin-PIN}
Y.P.Zhang, J.C.Liu, C.Guo, et~al.
\newblock The development of $^{222}{Rn}$ detectors for juno prototype.
\newblock {\em Radiation Detection Technology and Methods}, 2:5, 2018.

\bibitem{Wolfgang}
Wolfgang Rau.
\newblock Measurement of radon diffusion in polyethylene based on alpha
  detection.
\newblock {\em Nuclear Instruments and Methods in Physics Research A},
  664:65--70, 2012.

\end{thebibliography}
